\documentclass[journal]{IEEEtran}
\usepackage{xcolor}
\usepackage{cite}
\ifCLASSINFOpdf
  \usepackage[pdftex]{graphicx}
  \usepackage{xcolor}
  \graphicspath{{../figures/pdf/}{../figures/jpeg/}}
  \DeclareGraphicsExtensions{.pdf,.jpeg,.png}
\else
\fi
\usepackage{amsmath}
\usepackage{mathtools}

\usepackage{amssymb}
\usepackage{algorithm}
\usepackage{algpseudocode}

\usepackage{array}
\ifCLASSOPTIONcompsoc
 \usepackage[caption=false,font=normalsize,labelfont=sf,textfont=sf]{subfig}
\else
 \usepackage[caption=false,font=footnotesize]{subfig}
\fi
\usepackage{url}
\usepackage[protrusion=true,expansion=true]{microtype} % tighter justification, fewer overfull lines

\makeatletter
\g@addto@macro\normalsize{% IEEEtran resets display skips inside \normalsize; append after it
  \setlength{\abovedisplayskip}{5pt plus 1pt minus 0.5pt}%   default ~1.5ex (7.5pt)
}

\begin{document}
\title{Systematic SDP Search for Decentralized Stability Certificates}

\author{Andrey~Gorbunov,~\IEEEmembership{Member,~IEEE,}~Youhong~Chen,~\IEEEmembership{Member,~IEEE,}~Janusz~Bialek,~\IEEEmembership{Fellow,~IEEE,}, Balarko Chaudhuri,~\IEEEmembership{Fellow,~IEEE,},~Jin~Ma,~\IEEEmembership{Senior~Member,~IEEE,},~Gregor~Verbi\v{c},~\IEEEmembership{Senior~Member,~IEEE,} and Petr~Vorobev,~\IEEEmembership{Member,~IEEE}
\thanks{A. Gorbunov is with the Institute for Automation, ETH Zurich, Switzerland, and was previously with the School of Electrical and Computer Engineering, the University of Sydney, Australia. E-mail: agorbunov@ethz.ch.}
\thanks{Y. Chen is with the School of Electronic and Electrical Engineering, the University of Bath. E-mail: yc2930@bath.ac.uk}
\thanks{J. Bialek and B. Chaudhuri are with the Department of Electrical and Electronic Engineering, Imperial College London. E-mail: j.bialek@imperial.ac.uk, b.chaudhuri@imperial.ac.uk}
\thanks{J. Ma and G. Verbi\v{c} are with the School of Electrical and Computer Engineering, the University of Sydney, Australia. E-mail: j.ma@sydney.edu.au, gregor.verbic@sydney.edu.au}
\thanks{P. Vorobev is with the School of Electrical and Electronic Engineering, Nanyang Technological University. E-mail: petr.vorobev@ntu.edu.sg}

% % <-this % stops a space
% % % \thanks{Manuscript received April 19, 2005; revised August 26, 2015.}
 }


\markboth{REPLACE THIS LINE WITH YOUR MANUSCRIPT ID NUMBER}%
{Shell \MakeLowercase{\textit{et al.}}: Bare Demo of IEEEtran.cls for IEEE Journals}



% \IEEEaftertitletext{\vspace{-2\baselineskip}}
\maketitle

\begin{abstract}
Decentralized stability certificates allow inverter-based resources to be connected to a grid without repeating a system-wide small-signal study. Passivity provides such certificates but is violated by practical grid-forming (GFM) controls at low frequency, and its generalizations---transformation matrices, extended passivity indices, loop shifting, and gain--phase conditions---rely on multipliers that are constructed analytically or chosen heuristically. All of these certificates are linear matrix inequalities in their multipliers at every frequency, so the search for the least conservative multiplier is a convex semidefinite program (SDP) with a global optimum and a definite infeasibility verdict. This paper demonstrates how SDP can be applied to searching for optimal multipliers for some decentralized methods proposed in the literature. Then it illustrates the methodology by presenting a systematic search algorithm for the fully plug-and-play additive nodal-admittance certificate, which requires no assembled network model and holds for arbitrary interconnections. Only sampled frequency responses are needed, making the search compatible with black-box models and impedance scans. Chebyshev parameterization together with adaptive frequency splitting provides a piecewise-polynomial multiplier without manual partitioning. On the IEEE 39-bus system with ten droop-controlled GFMs, this multiplier certifies a larger region of droop gains than piecewise-constant multipliers, as confirmed by simulation of random inverter placements.

\end{abstract}


% \fi
\IEEEpeerreviewmaketitle

\section{Introduction}

\IEEEPARstart{P}{ower} systems worldwide are adopting inverter-based resources (IBRs) at a pace and scale without precedent. A growing body of operational evidence links high IBR penetration to oscillatory interactions spanning the sub- and super-synchronous ranges, driven by the coupling between converter controls and the electromagnetic dynamics of the network~\cite{chen2026subsynchronous}. Two features distinguish this problem from classical small-signal stability. First, converter dynamics are defined by control software rather than by electromechanical physics and extend over a wide range of time scales, so the quasi-static network models underpinning conventional analysis are no longer adequate. Second, the number of units is large: individual IBRs are far smaller than synchronous machines, and transmission systems already host thousands of converter-interfaced units where they previously hosted a few hundred generators.
 
These features expose two structural weaknesses of established small-signal tools. Modal analysis and the generalized Nyquist criterion deliver accurate results, but only for one network configuration, one set of connected devices and one operating point at a time; any change in topology, device placement or dispatch requires the assessment to be repeated, and a system-wide model of the required fidelity is rarely at the disposal of the system operator. In addition, vendors regard control algorithms as proprietary and typically disclose only black-box electromagnetic-transient (EMT) models, so a practically usable method must work with the terminal frequency response of a device---its small-signal impedance or admittance---rather than with its internal equations. Such frequency responses are routinely obtained from analytical models, from EMT models via frequency scans, or from measurements on hardware~\cite{harnefors2007input,sun2011impedance,wen2015small,cifuentes2022blackbox}.
 
Decentralized stability certification addresses both weaknesses~\cite{vorobev2019decentralized}: each device is required to satisfy a local condition on its terminal frequency response, chosen such that the interconnection of compliant devices through a suitably characterized network is guaranteed to be stable, without a system-wide model and without access to the internal control equations. The classical route to such conditions is passivity~\cite{Harnefors2016Passivity}. Since parallel and feedback interconnections of passive subsystems are passive, and hence stable, a network of passive lines populated by devices with passive terminal admittances is stable. Passivity has been used to derive decentralized stability conditions for power networks with dynamic lines~\cite{spanias2021passivity}, to establish transient stability of dispatchable virtual oscillator control~\cite{he2024passivity}, and as a design objective for converter control~\cite{huang2020hinf}. Its limitation is equally well known: realistic converter controls, including droop-controlled GFMs, are not passive at low frequencies, and no choice of droop gains removes this deficit~\cite{Harnefors2016Passivity,vorobev2019decentralized}. Strict passivity therefore rejects many designs that are stable in practice.
 
A considerable body of recent work relaxes passivity to reduce conservatism while retaining decentralized verification. \cite{laib2020decentralized} combines a bus-level passivity test with a second quadratic constraint on each bus together with its incident lines; \cite{dey2023passivity} exploits a slow--fast separation of the controller and a transformed low-frequency input--output description; \cite{niehues2025small} derives parametric conditions for voltage-droop control, and \cite{siahaan2024decentralized} proposes decentralized criteria for GFM control under lossless-line assumptions. \cite{haberle2026decentralized} compensates for the lack of passivity of the dynamic network by dynamic loop shifting and obtains parametric device-level certificates. \cite{chen2025extended} introduces frequency-dependent weighting matrices that define extended input-feedforward and output-feedback passivity indices for MIMO voltage-source converters. A complementary line of work replaces passivity with the small-phase theorem, based on the phases of sectorial complex matrices~\cite{chen2024phase}. \cite{huang2024gain} combines small-gain and small-phase conditions into decentralized certificates for systems containing GFMs, grid-following converters (GFLs) and synchronous machines, and \cite{woolcock2023mixed} derives a mixed gain/phase robustness criterion for structured perturbations. Very recent contributions address the failure of the sectoriality assumption for GFMs at low frequency through loop-shaping transformations~\cite{cifelli2025decentralized}. Geometric certificates based on the Davis--Wielandt shell have been unified in~\cite{feng2026unified}, where a shell separation is verified at each sampled frequency by an SDP in four scalar weights for a multiplier matrix fixed a priori.
 
Two observations motivate the present paper. First, despite their different derivations, all of these certificates take the same mathematical form, as also observed in~\cite{feng2026unified}: a positive-definiteness condition on the Hermitian part of a component frequency response, transformed by a weighting object---a transformation matrix~\cite{vorobev2019decentralized}, a loop-shifting operator~\cite{haberle2026decentralized}, a scaling matrix and rotation angle~\cite{huang2024gain}, or a weighting matrix and passivity indices~\cite{chen2025extended}. It is this weighting object, referred to in the following as the \emph{multiplier}, that governs the conservatism of the certificate. Second, in the existing literature, the multiplier is almost always constructed analytically or chosen heuristically; where it is optimized~\cite{feng2026unified}, this is done for a known network and one sampled frequency at a time. Analytical constructions exploit structural properties of the network or the controller and therefore require simplifying assumptions---lossless lines or a fixed $R/X$ ratio, neglected inner voltage and current loops, exact droop relations, or an explicit slow--fast separation---while heuristic choices, such as the rotation used in~\cite{vorobev2019decentralized}, are not guaranteed to exist for a given device and give no indication of how far from optimal they are.
 
In systems and control theory, by contrast, the search for multipliers is a classical convex problem. Frequency-domain inequalities that are linear in the multiplier constitute linear matrix inequalities (LMIs), and their solution by SDP is a mature technology~\cite{Boyd1994LMI,Vandenberghe1996SDP}. The integral quadratic constraint (IQC) framework parameterizes whole classes of multipliers and reduces robustness analysis to LMI feasibility~\cite{megretski1997iqc,veenman2016robust}; convex LMI searches have been developed for rational Zames--Falb multipliers~\cite{carrasco2014lmi}; and the Kalman--Yakubovich--Popov lemma converts frequency-domain inequalities for rational multipliers into finite-dimensional LMIs~\cite{rantzer1996kalman}. More recently, dissipativity properties have been verified directly from input--output data by convex optimization~\cite{romer2019oneshot}. In the power systems literature, systematic multiplier search has appeared only very recently, and only for certificates that compare a device against a known network~\cite{feng2026unified}. Our previous work parameterized a dynamic multiplier as a stable linear time-invariant (LTI) filter and tuned it by fixed-order structured $\mathcal{H}_\infty$ synthesis~\cite{gorbunov2027dynamic}; the resulting optimization is nonsmooth and nonconvex, provides no global optimality guarantee, and requires analytical transfer-function models.
\subsection{Main Contributions}
This paper turns the construction of decentralized stability certificates into a convex optimization problem that is solved from frequency-response data alone. The approach is not tied to one particular certificate: we give the precise SDP form of the multiplier search for each of the certificate families below, show how the frequency-dependent multiplier can be parameterized in a Chebyshev basis for any of them, and illustrate the complete construction on the additive nodal-admittance formulation of~\cite{vorobev2019decentralized}.

\subsubsection{Certificate search as a convex SDP}
Generalized passivity~\cite{vorobev2019decentralized}, extended passivity~\cite{chen2025extended}, dynamic loop shifting~\cite{haberle2026decentralized} and mixed gain--phase conditions~\cite{huang2024gain} are all LMIs in their multipliers at every frequency, and we state the corresponding SDP explicitly for each of them. Multiplier search at a given frequency is therefore an SDP: the optimum is global within the admitted class, and infeasibility proves that no multiplier of that class exists. Only sampled frequency responses are needed---from analytical models, linearized EMT models, impedance scans or measurements---and nothing is assumed about the controller structure.

\subsubsection{Automatic synthesis of frequency-dependent multipliers}
The Chebyshev parameterization applies to any of the certificates above; we illustrate it using the additive nodal-admittance formulation~\cite{vorobev2019decentralized}. The multiplier is parameterized in a Chebyshev basis, its coefficients are found by a single joint SDP, and a certificate-driven algorithm splits the frequency range wherever a low-degree local approximation is insufficient. The result is a piecewise-polynomial multiplier over the full frequency range, obtained without manual choice of frequency boundaries or sampling points and governed by two parameters with direct meaning: a certificate tolerance and a maximum local degree.

\section{SDP for Decentralized Frequency-Domain Stability }
\label{sec:sdp_certificates}

A broad class of decentralized frequency-domain stability certificates
for LTI systems can be expressed as positive-definiteness conditions involving the local
device frequency response, the network frequency response, and a set of
multiplier matrices. A generic local condition can be written as
\begin{equation}
\label{eq:generic_local_condition}
    L(\omega)
    + 2\operatorname{Her}\!\left(
        M(\omega)G(j\omega)
      \right)
    + G^{H}(j\omega)
        N(\omega)
        G(j\omega)
    \succ 0,
\end{equation}
where $\operatorname{Her}(X) :=
    \frac{1}{2}\left(X+X^{H}\right)$,
and $G$ may represent either a device admittance or impedance,
depending on the adopted feedback representation. The conditions \eqref{eq:generic_local_condition} are precisely the pointwise frequency form of IQC with multiplier $\Pi = \begin{bmatrix}
    L & M^H \\
    M & N
\end{bmatrix}$ \cite{megretski1997iqc}.
When the device and the network are interconnected in feedback, with the device
represented by its impedance $Z_{\mathrm{dev}}$ and the network by its admittance
$Y_{\mathrm{net}}$, the two subsystems are certified by \emph{complementary} IQCs conditions
\begin{align}
\label{eq:device_condition}
    L + 2\operatorname{Her}\!\left(M Z_{\mathrm{dev}}\right)
      + Z_{\mathrm{dev}}^{H} N Z_{\mathrm{dev}} &\succ 0, \\
\label{eq:network_condition}
    N + 2\operatorname{Her}\!\left(M^{H} Y_{\mathrm{net}}\right)
      + Y_{\mathrm{net}}^{H} L Y_{\mathrm{net}} &\preceq 0,
\end{align}
for all $\omega$: the network condition is obtained from the device condition by
interchanging $L \leftrightarrow N$, replacing $M$ by $M^{H}$ and
$Z_{\mathrm{dev}}$ by $Y_{\mathrm{net}}$, and reversing the sign.

The matrices $L$, $M$, and $N$ generalize conventional passivity.
% by defining a broader class of quadratic frequency-domain conditions.
For example, conventional strict passivity of \cite{Harnefors2016Passivity}, is recovered from
\eqref{eq:generic_local_condition} by choosing $(L, M, N) = (0, I, 0)$, which gives
\begin{equation}
\label{eq:standard_passivity}
    \operatorname{Her}\!\left(G(j\omega)\right)
    =
    \frac{1}{2}
    \left(
        G(j\omega)+G^{H}(j\omega)
    \right)
    \succ 0.
\end{equation}
Thus, conventional passivity corresponds to the identity multiplier.
More general multiplier matrices enlarge the admissible
frequency-domain region and can substantially reduce the conservatism
of \eqref{eq:standard_passivity}.

Fixed multipliers already extend conventional
passivity in the problematic frequency range \cite{vorobev2019decentralized, chen2025extended}. However, a single fixed
multiplier is often insufficient for practical IBR controls whose
frequency-domain characteristics vary substantially between the
low, medium, and high-frequency ranges. This motivates the use of
frequency-dependent, or \textit{dynamic}, multipliers. Importantly, the multiplier matrices $L$, $M$, $N$ enter conditions \eqref{eq:generic_local_condition} linearly. Hence, the resulting
positive-definite constraints constitute LMIs in the multiplier coefficients. Optimizing a convex function subject to these LMIs is an SDP \cite{Boyd1994LMI,Vandenberghe1996SDP}.

\section{SDP for Additive Nodal-Admittance Certificate} \label{sec:pnp}
First, let us consider the application of SDP to the additive nodal-admittance formulation approach presented in \cite{vorobev2019decentralized, gorbunov2027dynamic}. It derives decentralized stability certificates directly from system-level stability conditions through a homotopy argument, rather than using the passivity framework. Nevertheless, the resulting certificates admit an equivalent interpretation within the frequency-domain framework \eqref{eq:generic_local_condition} introduced in the previous section. 
 
\subsection{Decentralized Admittance Decomposition Method}
\label{subsec:no_crossing}
Let $N$ denote the total number of buses and
$v(s)$ the stacked vector of small-signal bus voltages in the conventional phasor representation.
Let $Y_{\mathrm{net}}$ denote the dynamic nodal
admittance of the interconnecting network and let
\begin{equation}
    Y_{\mathrm{dev}}(s)
    =
    \operatorname{blkdiag}
    \left(
        Y_1(s),\ldots,Y_N(s)
    \right)
\end{equation}
collect the aggregate device admittances at the corresponding buses.
Generators and loads are both included in this device representation,
with currents defined as flowing from the buses into the devices. $Y_i(s)=0_{2\times2}$ where no device is connected.

In the absence of external perturbations, the device-current vector
satisfies
\begin{equation}
    i_{\mathrm{dev}}(s)
    =
    -Y_{\mathrm{net}}(s)v(s)
    =
    Y_{\mathrm{dev}}(s)v(s).
\end{equation}
Combining the two equations above gives us $[Y_{\mathrm{net}}(s) + Y_{\mathrm{dev}}(s)] v(s) = 0$. The equation admits a nonzero voltage solution only if
\begin{equation}
\label{eq:characteristic_admittance}
    \det
    \left(
        Y_{\mathrm{net}}(s)+Y_{\mathrm{dev}}(s)
    \right)
    =0.
\end{equation}

Following \cite{vorobev2019decentralized}, introduce a continuous homotopy
\((Y_{\mathrm{net}}^\alpha,Y_{\mathrm{dev}}^\alpha)\), with
\(\alpha\in[0,1]\), connecting a known stable reference system
at \(\alpha=0\) to the system of interest at \(\alpha=1\).
Assume that the interconnection remains regular, that
the component admittances have no poles on the imaginary axis, and that
no modes enter through infinity along the homotopy.
Stability is then preserved provided that no mode crosses the imaginary
axis. A sufficient no-crossing condition is
\begin{equation}
\label{eq:no_crossing}
    \det
    \left(
        Y_{\mathrm{net}}^\alpha(j\omega)
        +
        Y_{\mathrm{dev}}^\alpha(j\omega)
    \right)
    \neq0
\end{equation}
for all \(\omega\geq0\) and all \(\alpha\in[0,1]\), where conjugate
symmetry of the real-rational \(dq\) models permits restriction to
nonnegative frequencies.

A convenient sufficient condition for
\eqref{eq:no_crossing} is
\begin{equation}
\label{eq:multiplier_global}
    \operatorname{Her}
    \left[
        M(\omega)
        \left(
            Y_{\mathrm{net}}^\alpha(j\omega)
            +
            Y_{\mathrm{dev}}^\alpha(j\omega)
        \right)
    \right]
    \succ 0,
\end{equation}
where $M(\omega)$ is nonsingular. Indeed, if the total admittance
were singular, then its product with any nonsingular $M$ would also be
singular and its Hermitian part could not be strictly positive
definite. This is the key observation used in
\cite{vorobev2019decentralized} to extend conventional passivity
through frequency-dependent multipliers. In addition, it could be shown that in the case of a multiplier $M(\omega)$ that does not depend on $\alpha$, it is necessary and sufficient to check positive definiteness only at endpoints $\alpha = 0$ and $\alpha = 1$ \cite{gorbunov2027dynamic}. In the following, we omit the $\alpha$ superscript.

\subsection{SDP-Based Certificates }
\label{subsec:pnp_certificate}
The main advantage of the admittance representation described above is its additive
structure. The total nodal admittance can be written as
\begin{equation}
\label{eq:additive_total_admittance}
    Y_{\mathrm{tot}}
    =
    Y_{\mathrm{dev}}
    +
    Y_{\mathrm{net}} ,
\end{equation}
Equivalently, $Y_{\mathrm{dev}}$ can be decomposed into a sum of
$N$ component matrices, where the $i$th component contains
$Y_{\mathrm{dev},i}$ in the $i$th $2\times2$ diagonal block and is
augmented by zeros in all remaining block positions.

Similarly, the network admittance can be decomposed into a sum of
individual line contributions. For a line connecting buses $i$ and
$j$, with dynamic $dq$ admittance
$Y_{\mathrm{line},ij}\in\mathbb{C}^{2\times2}$, the corresponding
component matrix is zero everywhere except in the four block positions
associated with buses $i$ and $j$. Its nonzero $2\times2$ block
submatrix is
\begin{equation}
\label{eq:Yline_component}
    \begin{bmatrix}
        Y_{\mathrm{line},ij}
        &
        -Y_{\mathrm{line},ij}
        \\
        -Y_{\mathrm{line},ij}
        &
        Y_{\mathrm{line},ij}
    \end{bmatrix}.
\end{equation}
Hence, both the device and network admittances are sums of local
component contributions. This additive structure is the key property
that enables component-wise decentralized stability conditions.

Consider a common block-diagonal multiplier
\begin{equation}
\label{eq:block_multiplier}
    M(\omega)
    =
    \operatorname{blkdiag}(m(\omega)),
    \qquad
    m(\omega)\in\mathbb{C}^{2\times2}.
\end{equation}
It could be shown \cite{vorobev2019decentralized} that using such a block-diagonal form of $M(\omega)$ allows to obtain component-wise sufficient stability conditions as follows:
\begin{equation}
\label{eq:local_device_passivity}
    \operatorname{Her}
    \left(
        m(\omega)
        Y_{\mathrm{dev},i}(j\omega)
    \right)
    \succ 0,
    \qquad \forall i,
\end{equation}
and
\begin{equation}
\label{eq:local_line_passivity}
    \operatorname{Her}
    \left(
        m(\omega)
        Y_{\mathrm{line},ij}(j\omega)
    \right)
    \succ 0,
    \qquad \forall (i,j)\in\mathcal{E},
\end{equation}
This construction provides a fully decentralized certificate: any device and any line satisfying the same local multiplier condition can be arbitrarily interconnected, and the resulting network is guaranteed to be stable regardless of its topology or size, much like the plug-and-play property of passivity.
These certificates depend on the operating point through the linearized device admittances. A common multiplier can also be sought by including several operating points as additional SDP constraints.

For a fixed $\omega$, the entries of the component admittances are
known complex numbers and $ \operatorname{Her}( mY) $
is linear in $m$. Hence, a multiplier
maximizing the smallest eigenvalue of $ \operatorname{Her}( mY)$ can be found from
\begin{subequations}
\label{eq:fixed_frequency_sdp}
\begin{align}
    \underset{m,\epsilon}{\operatorname{max}}
    \quad&
    \epsilon
    \label{eq:sdp_obj}
    \\
    \operatorname{s.t.}
    \quad&
    \operatorname{Her}
    \left(
        mY_{\mathrm{dev},i}(j\omega)
    \right)
    \succeq
    \epsilon I_2,
    \quad \forall i,
    \label{eq:sdp_device}
    \\
    &
    \operatorname{Her}
    \left(
        mY_{\mathrm{line},ij}(j\omega)
    \right)
    \succeq
    \epsilon I_2,
    \quad \forall (i,j)\in\mathcal{E},
    \label{eq:sdp_line}
    \\
    &
    \| m \| \leq 1.
    \label{eq:sdp_normalization}
\end{align}
\end{subequations}
Since the minimum eigenvalue $\epsilon$ of a linear Hermitian
matrix is concave, \eqref{eq:fixed_frequency_sdp} is a convex SDP \cite{Vandenberghe1996SDP}.

If $\epsilon^\star>0$, the resulting multiplier gives a strict
certificate at the selected frequency. Moreover, because at least one
nonsingular component satisfies a strictly positive-definite
transformed condition, the resulting multiplier is necessarily
nonsingular. The optimized margin also provides tolerance to operating-point variations and other admittance uncertainties. Specifically, if the actual admittance is $Y+E$, with $\|E\|_2\le \epsilon$, its transformed Hermitian part remains positive definite whenever $\operatorname{Her}(mY) \succeq \epsilon I_2$ given $\|m\|_2\leq 1$. Thus, operating-point changes covered by uncertainty $E$ preserve the certificate without repeating the multiplier search. Alternatively, admittances at several operating points can be included directly as additional SDP constraints.

Therefore, enforcing the constraints on a prescribed frequency set, $\{\omega_1, \dots, \omega_K\}$, leads directly to an SDP. The issue of guaranteeing the
inequalities over the complete continuous frequency interval, rather
than only at sampled points, is considered separately in
Section~\ref{sec:continuous_frequency_guarantees}.

\section{SDP Applied to Other Passivity-Based Decentralized Methodologies}
\label{sec:other_conditions}

The multiplier formulation $MY$ presented in the previous section is particularly attractive because
of the additive structure of nodal admittances resulting in fully topology agnostic stability conditions. In this section, we explicitly demonstrate how several representative decentralized stability conditions can be written as an SDP. Due to space limitations, the complete decentralized formulations are omitted. These additionally require suitable structural constraints on the weighting matrices---for example, block-diagonal, repeated-block, or sparsity patterns that preserve decomposition into local device-level tests---as well as appropriate normalization.

\subsection{Extended frequency-domain passivity}
A closely related extension of conventional passivity was proposed in
\cite{chen2025extended}. For a MIMO transfer matrix $G$, a
frequency-dependent weighting matrix $R(\omega)$ is introduced and the
extended input-feedforward and output-feedback passivity indices are
defined from
\begin{align}
    \operatorname{Her}
    \left(
        R^{H}(\omega)G(j\omega)
    \right),
    \label{eq:chen_ifp}
    \\
    \operatorname{Her}
    \left(
        R^{H}(\omega)G^{-1}(j\omega)
    \right).
    \label{eq:chen_ofp}
\end{align}
For two interconnected systems $G_1$ and $G_2$, the corresponding
stability condition can be represented by the existence of scalars
$\nu(\omega)$ and $\rho(\omega)$ satisfying
\begin{align}
    \operatorname{Her}
    \left(
        R^{H}G_1
    \right)
    -
    \nu I
    &\succeq0,
    \label{eq:chen_lmi_1}
    \\
    \operatorname{Her}
    \left(
        R^{H}G_2^{-1}
    \right)
    -
    \rho I
    &\succeq0,
    \label{eq:chen_lmi_2}
    \\
    \nu+\rho &>0.
    \label{eq:chen_lmi_sum}
\end{align}
For fixed frequency responses,
\eqref{eq:chen_lmi_1}--\eqref{eq:chen_lmi_sum} are linear in
$R$, $\nu$, and $\rho$. These conditions can therefore also be incorporated
into an SDP-based search. Noteworthy, \cite{chen2025extended}
also suggested that convex optimization could be used to search for
the multiplier $R(\omega)$, although a detailed implementation was
not provided. 

\subsection{Dynamic Loop Shifting}
\label{subsubsec:loop_shifting}

A related approach for relaxing conventional passivity conditions is dynamic loop shifting. This approach is employed in \cite{haberle2026decentralized} to compensate for the lack of passivity of dynamic power-network models and to derive decentralized stability certificates.

Consider the negative-feedback interconnection of two transfer
matrices $Z_{\mathrm{dev}}(s)$ and $Y_{\mathrm{net}}(s)$. Following
\cite{haberle2026decentralized}, introduce a dynamic transfer matrix
$\Gamma(s)$ as a feedforward around $Y_{\mathrm{net}}(s)$ and a feedback around
$Z_{\mathrm{dev}}(s)$. The corresponding loop-shifted subsystems are
\begin{align}
\label{eq:loop_shift_Zdev}
    Z_{\mathrm{dev}}'(s)
    &=
    \left(
        I-Z_{\mathrm{dev}}(s)\Gamma(s)
    \right)^{-1}
    Z_{\mathrm{dev}}(s),
    \\
\label{eq:loop_shift_Ynet}
    Y_{\mathrm{net}}'(s)
    &=
    Y_{\mathrm{net}}(s)+\Gamma(s).
\end{align}

Under the regularity assumptions stated in
\cite{haberle2026decentralized}, stability of the loop-shifted
interconnection $Z_{\mathrm{dev}}'\#Y_{\mathrm{net}}'$ implies stability of the original
interconnection $Z_{\mathrm{dev}}\#Y_{\mathrm{net}}$.

Although the loop-shifting transformation itself is expressed
in terms of the impedance--admittance feedback pair
\eqref{eq:loop_shift_Zdev}--\eqref{eq:loop_shift_Ynet}, its passivity
conditions can be written as the pair of positive-semidefinite
inequalities
\begin{align}
\label{eq:loop_shift_affine_device}
    \operatorname{Her}
    \left(
        Y_{\mathrm{dev}}(j\omega)-\Gamma(j\omega)
    \right)
    &\succ0,
    \\
\label{eq:loop_shift_affine_network}
    \operatorname{Her}
    \left(
        Y_{\mathrm{net}}(j\omega)+\Gamma(j\omega)
    \right)
    &\succeq0.
\end{align}
This representation corresponds to \eqref{eq:generic_local_condition} with $L = \pm \frac{1}{2} (\Gamma + \Gamma^H)$, $M = I$, and $N = 0$. It is particularly useful from an optimization
perspective because $\Gamma(j\omega)$ enters both conditions
\emph{linearly}.

At a fixed frequency, a loop-shifting transfer matrix maximizing a
common passivity margin can therefore be searched from
\begin{subequations}
\label{eq:loop_shift_sdp}
\begin{align}
    \underset{\Gamma,\,\epsilon}{\operatorname{max}}
    \quad&
    \epsilon
    \\
    \operatorname{s.t.}
    \quad&
    \operatorname{Her}
    \left(
        Y_{\mathrm{dev}}(j\omega)
        -
        \Gamma
    \right)
    \succeq
    \epsilon I,
    \label{eq:loop_shift_sdp_device}
    \\
    &
    \operatorname{Her}
    \left(
        Y_{\mathrm{net}}(j\omega)
        +
        \Gamma
    \right)
    \succeq
    \epsilon I ,
    \label{eq:loop_shift_sdp_network}
\end{align}
\end{subequations}
together with structural constraints on $\Gamma$, such as a block-diagonal form required for decentralized device verification. Since the objective is linear and the two matrix inequalities
are linear in $\Gamma$ and in
$\epsilon$, \eqref{eq:loop_shift_sdp} is again an SDP
\cite{Boyd1994LMI,Vandenberghe1996SDP}.

In \cite{haberle2026decentralized},  block-diagonal $\Gamma(s)$ is obtained
analytically using structural properties of the network dynamics so
that the loop-shifted network becomes passive. The observation above
shows that, when $Y_{\mathrm{dev}}(j\omega)$, $Y_{\mathrm{net}}(j\omega)$ data are available, the same
loop-shifting principle naturally suggests an alternative
optimization-based search for $\Gamma$.

\subsection{Small-Gain Conditions}
\label{subsubsec:small_gain}

Consider the feedback interconnection between the network admittance
$Y_{\mathrm{net}}(s)$ and the block-diagonal device impedance
$Z_{\mathrm{dev}}(s)$. A
sufficient decentralized conditions  based on the small-gain theorem derived in \cite{huang2024gain} are the following:
\begin{equation}
\label{eq:small_gain_scaled}
    \left\|
        DZ_{\mathrm{dev}}
    \right\|_2<1,
    \qquad
    \left\|
        Y_{\mathrm{net}}D^{-1}
    \right\|_2<1 ,
\end{equation}
where $D \succ 0$ is a nonsingular scaling matrix.

Let $ P:=D^{H}D$. The first inequality in \eqref{eq:small_gain_scaled} is equivalent to
\begin{equation}
\label{eq:sg_device}
    I
    -
    Z_{\mathrm{dev}}^{H}(j\omega)
    P(\omega)
    Z_{\mathrm{dev}}(j\omega)
    \succ0.
\end{equation}
Similarly, the second inequality is equivalent, after a congruence transformation by $D$, to
\begin{equation}
\label{eq:sg_network}
    P(\omega)
    -
    Y_{\mathrm{net}}^{H}(j\omega)
    Y_{\mathrm{net}}(j\omega)
    \succ0.
\end{equation}
Both \eqref{eq:sg_device}--\eqref{eq:sg_network} are LMIs in matrix $P(\omega)$ and can therefore be incorporated directly into following SDP:

\begin{subequations}
\label{eq:small_gain_sdp}
\begin{align}
    \underset{P,\,\epsilon}{\operatorname{max}}
    \quad&
    \epsilon
    \\
    \operatorname{s.t.}
    \quad&
    I
    -
    Z_{\mathrm{dev}}^{H}(j\omega)
    P(\omega)
    Z_{\mathrm{dev}}(j\omega)
    \succ\epsilon I,
    \label{eq:small_gain_sdp_device}
    \\
    &
    P(\omega)
    -
    Y_{\mathrm{net}}^{H}(j\omega)
    Y_{\mathrm{net}}(j\omega)
    \succ\epsilon I.
    \label{eq:small_gain_sdp_network}
\end{align}
\end{subequations}

\subsection{Small-Phase Conditions}
\label{subsubsec:small_phase}

Small-phase conditions provide an alternative characterization based
on the phases of complex matrices \cite{huang2024gain}. Particularly, small-phase conditions are that there exists a rotation angle $\theta$ such that the
two appropriately rotated numerical ranges lie in the right half-plane,
\begin{equation}
\label{eq:phase_device}
    \operatorname{Her}
    \left(
        D e^{-j\theta}
        Z_{\mathrm{dev}}(j\omega)
    \right)
    \succ0 , 
    \\
    \operatorname{Her}
    \left(
        D e^{j\theta}
        Y_{\mathrm{net}}
    \right)
    \succ0.
\end{equation}

The weight matrices $M_{\mathrm{dev}} = D e^{-j\theta}$, $M_{\mathrm{net}} = D e^{j\theta}$ can therefore be interpreted as particular multiplier matrices. This observation suggests that small-phase conditions are the
specific parameterization of weight matrix $M = D e^{\pm j\theta}$ in decentralized certification \eqref{eq:multiplier_global}.

\subsection{Generalized Gain--Phase Conditions}
\label{subsubsec:generalized_gain_phase}

The small-gain and small-phase conditions above can be embedded in a
common family of positive-definiteness constraints \eqref{eq:generic_local_condition}. For the device--network
interconnection, consider the complementary conditions
\begin{align}
\label{eq:generalized_device}
    L
    + 2\operatorname{Her}\!\left(
        MZ_{\mathrm{dev}}
      \right)
    + Z^{H}_{\mathrm{dev}}
        N
        Z_{\mathrm{dev}}
    \succ 0,
    \\
\label{eq:generalized_network}
    -N
    +
    2 \operatorname{Her}(M^{H}Y_{\mathrm{net}})
    -Y_{\mathrm{net}}^{H}LY_{\mathrm{net}}
    \succ0.
\end{align}

Both small-gain and small-phase conditions correspond to particular
choices of $L$, $M$, and $N$, summarized in
Table~\ref{tab:gain_phase_multipliers}.
\begin{table}[t]
\centering
\caption{Small-gain and small-phase parameterizations}
\label{tab:gain_phase_multipliers}
\renewcommand{\arraystretch}{1.1}
\begin{tabular}{c|c|c}
\hline
 & Small gain & Small phase \\
\hline
Device
&
$(L,M,N)=(I,0,-D^2)$
&
$(L,M,N)=(0,De^{-j\theta},0)$
\\
Network
&
$(L,M,N)=(D^2,0,-I)$
&
$(L,M,N)=(0,De^{j\theta},0)$
\\
\hline
\end{tabular}
\end{table}

The common representation also permits gain- and phase-type
certificates to be combined. For example, for nonnegative weights
$\lambda_{\mathrm{g}}$ and $\lambda_{\mathrm{p}}$, the choice
\begin{equation}
\label{eq:weighted_mixed_device}
    L
    =
    \lambda_{\mathrm{g}}I,
    \qquad
    M
    =
    \lambda_{\mathrm{p}}De^{-j\theta},
    \qquad
    N
    =
    -\lambda_{\mathrm{g}}P
\end{equation}
produces a weighted combination of the gain- and
phase-type conditions. 
Hence, the pure small-gain and small-phase certificates are recovered
by setting, respectively,
$\lambda_{\mathrm{p}}=0$ or $\lambda_{\mathrm{g}}=0$, while nonzero
values of both weights provide a mixed small gain-phase certificate.

More generally, rather than prescribing the gain- or phase-type
structures in Table~\ref{tab:gain_phase_multipliers}, the matrices
$L$, $M$, and $N$ can themselves be treated as decision variables.
For fixed $Y_{\mathrm{net}}(j\omega)$ and
$Z_{\mathrm{dev}}(j\omega)$, conditions
\eqref{eq:generalized_device}--\eqref{eq:generalized_network}
are linear in $L$, $M$, and $N$. Therefore, after imposing suitable
normalization and structural constraints on weight matrices, the search for a generalized
certificate at a given frequency can be formulated as SDP:
\begin{subequations}
\label{eq:mixed_small_gain_sdp}
\begin{align} 
    \underset{L,M,N,\,\epsilon}{\operatorname{max}}
    \quad&
    \epsilon
    \\
    \operatorname{s.t.}
    \quad&
    L
    +
    2\operatorname{Her}(MZ_{\mathrm{dev}})
    +
    Z_{\mathrm{dev}}^{H}NZ_{\mathrm{dev}} \succ \epsilon I,
    \label{eq:mixed_small_gain_sdp_device}
    \\
    &
    -N
    + 2\operatorname{Her}(M^{H}Y_{\mathrm{net}})
    -Y_{\mathrm{net}}^{H}LY_{\mathrm{net}}
    \succ\epsilon I.
    \label{eq:mixed_small_gain_sdp_network}
\end{align}
\end{subequations}
The DW-shell criterion of~\cite{feng2026unified}, with four nonnegative weights for a fixed multiplier matrix, is a member of this family.

We proceed to illustrate the details of the SDP-based certificate search using the additive
component-wise conditions
\eqref{eq:local_device_passivity}--\eqref{eq:local_line_passivity} for droop-controlled GFM-based systems. The described SDP-based synthesis procedure applies to any general conditions described by \eqref{eq:mixed_small_gain_sdp}.

\section{Search Algorithm for Frequency-Dependent Multipliers}
\label{sec:continuous_frequency_guarantees}

The fixed-frequency SDP in
\eqref{eq:fixed_frequency_sdp} provides a systematic way of finding a
multiplier at an individual frequency. For a plug-and-play certificate,
however, $m(j\omega)$ must be specified over the complete frequency
range of interest.

Analytical constructions \cite{haberle2026decentralized,huang2024gain} and rational parameterizations via the KYP lemma \cite{rantzer1996kalman} and structured $\mathcal{H}_{\infty}$ synthesis \cite{gorbunov2027dynamic} require explicit transfer-function models and are either restrictive or nonconvex. We instead retain the frequency-wise SDP and construct $m(j\omega)$ directly from sampled responses, which may originate from analytical models, linearizations, impedance scans, or measurements. The SDP-based certificates of~\cite{feng2026unified} are likewise verified only on a manually chosen frequency grid, without a guarantee between the sampled points.

\subsection{Piecewise-Constant Multipliers}
\label{subsec:piecewise_const}

A simple extension of a fixed multiplier is a piecewise-constant
choice
\begin{equation}
    m(j\omega)=m_r,
    \qquad
    \omega\in\mathcal{I}_r ,
\end{equation}
where $\mathcal{I}_r$ denotes a prescribed frequency region. This
already exploits the observation that different transformations may
be suitable at low and high frequencies. For example, conventional
passivity can be retained at high frequency using $m=I_2$, while a heuristic $m = \begin{bmatrix}
    0 & -1 \\ 1 & 0
\end{bmatrix}$ rotation is applied in the low-frequency region where
IBR admittances commonly violate passivity
\cite{vorobev2019decentralized,chen2025extended}. Such a heuristic
certifies stability for some parameter ranges but remains conservative.

To reduce this conservatism, the multiplier for a given region need not
be fixed a priori. For a single region
$\mathcal{I}_r$, a constant multiplier $m_r$ can
instead be computed directly:
\begin{subequations}
\label{eq:piecewise_sdp}
\begin{align}
    \underset{m,\,\{\epsilon_i\}}{\operatorname{max}}
    \quad&
    \sum_{i=1}^{N}\epsilon_i
    \\
    \operatorname{s.t.}
    \quad&
    \operatorname{Her}
    \left(
        mY_{\mathrm{dev}}(j\omega_i)
    \right)
    \succeq
    \epsilon_i I_2,
    \qquad
    \omega_i\in\mathcal{I},
    \\
    &
    \operatorname{Her}
    \left(
        mY_{\mathrm{line}}(j\omega_i)
    \right)
    \succeq
    \epsilon_i I_2,
    \qquad
    \omega_i\in\mathcal{I},
    \\
    &
    \|m\|_F\leq1 .
\end{align}
\end{subequations}
The summed margin is used here because the transformed GFM admittance
can become singular at $\omega=0$, making a uniform minimum-margin
objective unnecessarily restrictive.

\begin{figure}[t]
    \centering
    \subfloat[]{\includegraphics[width=0.7\linewidth]{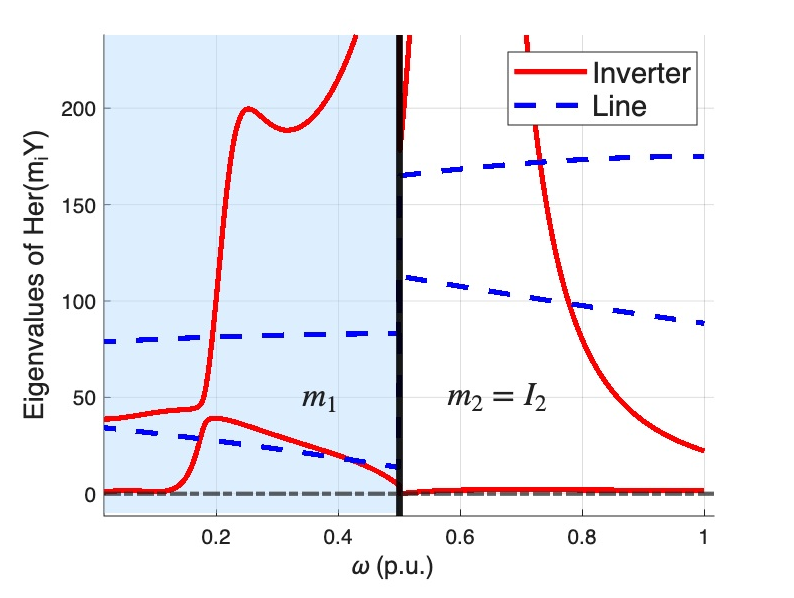}%
        \label{fig:piecewise_single}}\\
    \subfloat[]{\includegraphics[width=0.7\linewidth]{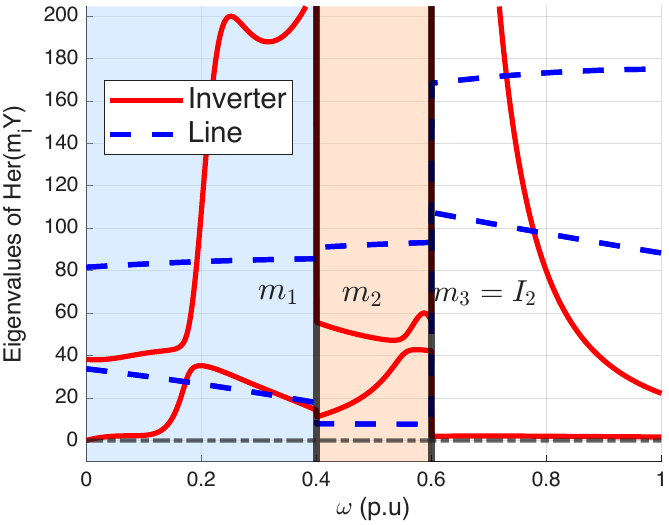}%
        \label{fig:three_regions_split}}
    \caption{Eigenvalues of the transformed Hermitian parts of the GFM
    admittance (solid red) and line admittance (dashed blue) for
    piecewise-constant multipliers: (a) two frequency regions, with $m_1$
    from \eqref{eq:piecewise_single} at low frequency and $m_2=I_2$ at
    high frequency; (b) three frequency regions, with the multipliers
    of \eqref{eq:m_SDP_3}.}
    \label{fig:piecewise_constant}
\end{figure}

We now show some examples of how the constant-piecewise multiplier
matrices can be derived using \eqref{eq:piecewise_sdp} before introducing
the global Chebyshev multiplier search. The certificate is obtained for
the detailed GFM converter model of \cite{pogaku2007modeling},
with the exact parameter values used given in the Appendix.
Solving \eqref{eq:piecewise_sdp} in the low-frequency range
$\omega \in (0, 0.5)$ (p.u.) with $1000$ samples gives
\begin{equation}
\label{eq:piecewise_single}
m_1 =
\begin{bmatrix}
0.17 - 0.032j & -0.69 - 0.12j \\
0.60 - 0.36j &  0.00
\end{bmatrix}.
\end{equation}
Together with $m_2=I_2$ at higher frequencies, this provides the
two-region certificate shown in Fig.~\ref{fig:piecewise_single}.

Splitting the frequency into more regions can reduce conservativeness further.
For example, using three manually selected regions ($[0, 0.4]$ and $[0.4, 0.6]$) produces the following
multipliers 
\begin{subequations}
\label{eq:m_SDP_3}
\begin{align}
m_1 &=
\begin{bmatrix}
0.21 - 0.04j &  -0.67 - 0.15j \\
0.57 - 0.39j &   0.00
\end{bmatrix},
\\
m_2 &=
\begin{bmatrix}
0.058 + 0.48j &  -0.55 + 0.03j \\
0.49 - 0.01j &  0.06 + 0.48j
\end{bmatrix} .
\end{align}
\end{subequations}
The corresponding transformed admittances are shown in
Fig.~\ref{fig:three_regions_split}.
Although effective, this procedure still requires manual choices of
the frequency boundaries and sampling frequencies. Moreover, a
piecewise-constant multiplier does not exploit the full flexibility
allowed by the frequency-dependent certificate. We therefore seek a
systematic representation of $m(j\omega)$ whose coefficients can be
optimized directly using SDP.

\subsection{Global Chebyshev Multiplier Search}
\label{subsec:global_cheb}
Consider first the complete frequency range
$\Omega = [\omega_{\min},\omega_{\max}]$ and map it to $x\in[-1,1]$
according to
\begin{equation}
\label{eq:global_frequency_mapping}
    x(\omega)
    =
    \frac{
        2\omega-\omega_{\min}-\omega_{\max}
    }{
        \omega_{\max}-\omega_{\min}
    }.
\end{equation}
The component admittances are approximated as
\begin{equation}
\label{eq:global_cheb_Y}
    \widehat Y_\ell^{(d)}(j\omega)
    =
    \sum_{k=0}^{d}
        A_{\ell,k}\,T_k\big(x(\omega)\big),
\end{equation}
where $T_k$ are the Chebyshev polynomials of the first kind and the
coefficients $A_{\ell,k}$ are obtained by interpolating queried
frequency responses at the $d+1$ Chebyshev--Lobatto frequencies
\begin{equation}
\label{eq:global_cheb_nodes}
    \omega_\nu
    =
    \frac{\omega_{\min}+\omega_{\max}}{2}
    +
    \frac{\omega_{\max}-\omega_{\min}}{2}
    \cos\left(\frac{\nu\pi}{d}\right) .
\end{equation}
These frequencies are not spaced uniformly but are placed more densely
towards the edges of the band, which is known from approximation
theory to be close to the best possible choice of sample points:
interpolation at these nodes is numerically well conditioned, avoids
the strong endpoint oscillations of equidistant high-order polynomial
interpolation, and gives an approximation $\widehat Y_\ell^{(d)}$
that is nearly as accurate as the best polynomial of the same degree,
with an error that decreases exponentially in $d$ for smooth
admittances \cite{trefethen2019approximation}.

The multiplier is parameterized in the same basis,
\begin{equation}
\label{eq:global_cheb_multiplier}
    m^{(d)}(j\omega)
    =
    \sum_{k=0}^{d}
        C_kT_k(x(\omega)),
\end{equation}
where $C_k$ are optimization variables. At the $d+1$ interpolation
frequencies, a degree-$d$ multiplier has the same pointwise freedom as
$d+1$ independent matrices $m(j\omega_\nu)$. Hence, instead of
solving independent SDPs and subsequently fitting their solutions, the
Chebyshev coefficients can be found directly from one joint SDP:
\begin{subequations}
\label{eq:global_cheb_sdp}
\begin{align} \label{eq:SDP_Chebyshev}
    \underset{\{C_k\},\,\epsilon}
    {\operatorname{max}}
    &
    \epsilon 
    \\ \label{eq:SDP_Chebyshev_device}
    \operatorname{s.t.}
    &
    \operatorname{Her}
    \left[
        \sum_{k=0}^d C_k T_k(x(\omega_\nu))
        Y_\ell(j\omega_\nu)
    \right]
    \succeq
    \epsilon I_2, 
    \forall \nu, \ell,
    \\
    &
    \sum_{k=0}^{d}\|C_k\|_F\leq1 .
\end{align}
\end{subequations}
The admittance values in
\eqref{eq:global_cheb_sdp} are fixed data and
$m^{(d)}(j\omega_\nu)$ is affine in $C_k$; therefore the optimization
remains an SDP.

In principle, this construction \eqref{eq:global_cheb_sdp} is already sufficient. If
$\widehat Y_\ell^{(d)}$ approximates $Y_\ell$ accurately over
$\Omega$, then $\operatorname{Her}
    [
        m
        \widehat Y_\ell^{(d)}
    ]$
provides an accurate approximation of
$\operatorname{Her}
    [
        m
        Y_\ell
    ]$.
Indeed,
\begin{equation}
\label{eq:H_error_from_Y}
    \|
        \operatorname{Her}[
        m
        \widehat Y_\ell^{(d)}]-\operatorname{Her}[
        m Y_\ell]
    \|_2
    \leq
    \|m\|_2
    \|
        Y_\ell-\widehat Y_\ell^{(d)}
    \|_2 .
\end{equation}

\begin{figure}[t]
    \centering
    \includegraphics[width=0.88\linewidth]{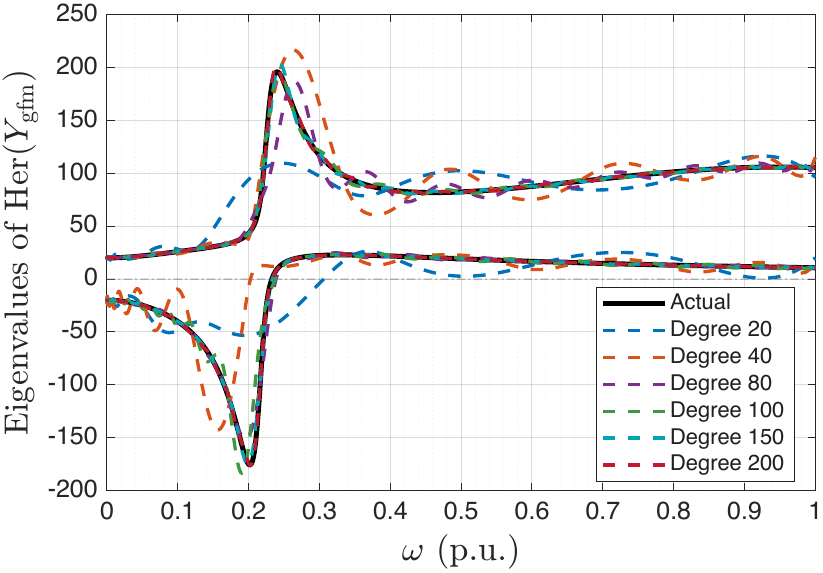}
    \caption{Hermitian part of the GFM admittance obtained from the
    original frequency response $Y_{\mathrm{inv}}(j\omega)$ and from
    its global Chebyshev approximation
    $\widehat Y_{\mathrm{inv}}^{(d)}(j\omega)$ for different
    polynomial degrees $d$.}
    \label{fig:gfm_global_cheb_accuracy}
\end{figure}

However, for practical GFM admittance, polynomial degrees greater than $d=200$ are required before the global approximation of $Y$ reaches a relative Frobenius error of $1\%$. This is reflected in Fig.~\ref{fig:gfm_global_cheb_accuracy}, which compares the eigenvalues of $\operatorname{Her}(Y)$ obtained from the global approximation with those of the exact admittance for increasing degree.

Such a high degree is undesirable for the multiplier search. For a
full complex $2\times2$ multiplier,
\eqref{eq:global_cheb_multiplier} introduces
\begin{equation}
\label{eq:global_variable_count}
    n_{\mathrm{var}}
    =
    8(d+1)+1
\end{equation}
real optimization variables, including the certificate margin. Thus,
$d=200$ already corresponds to more than $1600$ direct multiplier
variables. Moreover, when $K$ component admittances are certified
simultaneously, the $d+1$ interpolation frequencies generate $N_{\mathrm{LMI}} = K(d+1)$
positive-semidefinite constraints. Although each local constraint is
only $2\times2$, a very high global polynomial order unnecessarily
increases both computational burden and numerical sensitivity. This motivates replacing a single global high-degree approximation by several low-degree local approximations.

Before we switch to adaptive frequency splitting, we note that the parameterization~\eqref{eq:global_cheb_multiplier} extends directly to the other decentralized certificates in the class of quadratically constrained problems~\eqref{eq:mixed_small_gain_sdp} of Section~\ref{sec:other_conditions}. Expanding each multiplier in the same basis,
\begin{subequations}\label{eq:global_cheb_multiplier_general}
    \begin{equation}
        N^{(d)}(j\omega) = \sum_{k=0}^{d} D_k,T_k(x(\omega)),
    \end{equation}
    \begin{equation}
        L^{(d)}(j\omega) = \sum_{k=0}^{d} F_k,T_k(x(\omega)),
    \end{equation}
\end{subequations}
with coefficient matrices $D_k$ and $F_k$ as optimization variables, and substituting into~\eqref{eq:mixed_small_gain_sdp} at the interpolation frequencies $\omega_\nu$ produces a single joint SDP of the same form as~\eqref{eq:global_cheb_sdp}: the frequency responses are fixed data, the constraints are linear in $\{C_k, D_k, F_k\}$.

\subsection{Adaptive Frequency Splitting}
\label{subsec:adaptive_split}

Let the complete range be partitioned as
\begin{equation}
\label{eq:adaptive_partition}
    [\omega_{\min},\omega_{\max}]
    =
    \bigcup_{r=1}^{N_{\mathrm{reg}}}
    \mathcal{I}_r ,
    \qquad
    \mathcal{I}_r
    =
    [\omega_r^{-},\omega_r^{+}].
\end{equation}
On every interval, map $[\omega_r^{-},\omega_r^{+}]$ to $x_r \in [-1, 1]$ according to \eqref{eq:global_frequency_mapping} and construct local approximations $\widehat Y_{\ell,r}^{(d)}(j\omega)$ and $m_r^{(d)}(j\omega)$.
The multiplier coefficients are again found from the sampled-frequency
SDP \eqref{eq:global_cheb_sdp}, applied locally to
$\mathcal{I}_r$.

The central idea is to keep the local polynomial degree moderate and
split an interval whenever the approximation is insufficient. The
refinement is driven by the certificate itself rather than by the
admittance approximation alone. Define
\begin{align}
    H_{\ell,r}(\omega)
    &=
    \operatorname{Her}
    \left[
        m_r(j\omega)Y_\ell(j\omega)
    \right],
    \\
    \widehat H_{\ell,r}(\omega)
    &=
    \operatorname{Her}
    \left[
        m_r(j\omega)\widehat Y_{\ell,r}(j\omega)
    \right].
\end{align}
At independent validation frequencies, the certificate discrepancy is
measured as
\begin{equation}
\label{eq:certificate_error_split}
    \widehat e_{H,r}
    =
    \max_{\ell}
    \max_{\omega\in\Omega_r^{\mathrm{val}}}
    \left\|
        H_{\ell,r}(\omega)
        -
        \widehat H_{\ell,r}(\omega)
    \right\|_2 .
\end{equation}
The interval is considered sufficiently resolved when
$\widehat e_{H,r}
    \leq e_H $
and the directly evaluated transformed admittances remain positive
over the validation set. If this condition cannot be achieved with
$d\leq d_{\max}$, the interval is divided and the procedure is
repeated locally. The split point is chosen at the validation frequency where the smallest eigenvalue of $H_{\ell,r}$ is minimal if the positivity check fails, and otherwise where the certificate discrepancy is largest. To avoid excessively small subintervals, the split point is clipped to the central $60\%$ of the interval.

The resulting multiplier is piecewise polynomial,
\begin{equation}
\label{eq:final_piecewise_cheb}
    m(j\omega)
    =
    m_r(j\omega),
    \qquad
    \omega\in\mathcal{I}_r.
\end{equation}
Unlike the piecewise-constant construction of
Section~\ref{subsec:piecewise_const}, both the frequency partition and
the frequency dependence within each region are determined automatically.

\begin{algorithm}[t]
\caption{Adaptive multiplier synthesis}
\label{alg:turnkey_multiplier}
\begin{algorithmic}[1]

\Require
certificate tolerance $e_H$,
maximum degree $d_{\max}$.

\State Initialize the whole frequency range as one interval.

\While{unresolved intervals remain}
    \State Solve SDP \eqref{eq:SDP_Chebyshev} for $m^{(d)}_r(j\omega)$.

    \State Validate
    $\widehat e_{H,r}$
    at additional frequencies \eqref{eq:certificate_error_split}.

    \If{$\widehat e_{H,r} \leq e_H$}
        \State Accept the interval.
    \ElsIf{$\min_\omega H_{\ell,r}<0$}
        \State Split the interval at the smallest $H_{\ell,r}$
    \Else
        \State Split the interval at the largest $\|H_{\ell,r} - \widehat{H}_{\ell,r} \|_2$
    \EndIf
\EndWhile

\State Assemble the accepted local multipliers according to
\eqref{eq:final_piecewise_cheb}.

\end{algorithmic}
\end{algorithm}

The two user-defined numerical parameters therefore have direct
interpretations. The tolerance $e_H$ specifies the required accuracy
of the quantity entering the stability certificate, while
$d_{\max}$ limits the size of each local SDP. The latter can also be
specified indirectly through a maximum number of optimization
variables using \eqref{eq:global_variable_count} for an unrestricted complex $2\times2$ multiplier. Adaptive splitting
then trades a larger number of small SDPs against a single
high-dimensional global optimization.

The continuous-frequency certification is supported by the convergence properties of Chebyshev interpolation: for admittances analytic in a neighbourhood of each frequency interval, the approximation converges uniformly at a geometric rate \cite{trefethen2019approximation}. Through \eqref{eq:H_error_from_Y}, this also controls the approximation error in the transformed certificate. At the finite degrees used in practice, the error is assessed numerically through adaptive refinement and independent validation. Additional information about the admittance functions, such as pole locations and magnitude bounds, would allow explicit analytical error bounds.

\section{Case Study} \label{sec:case}

\begin{figure}
    \centering
    \includegraphics[width=0.7\linewidth]{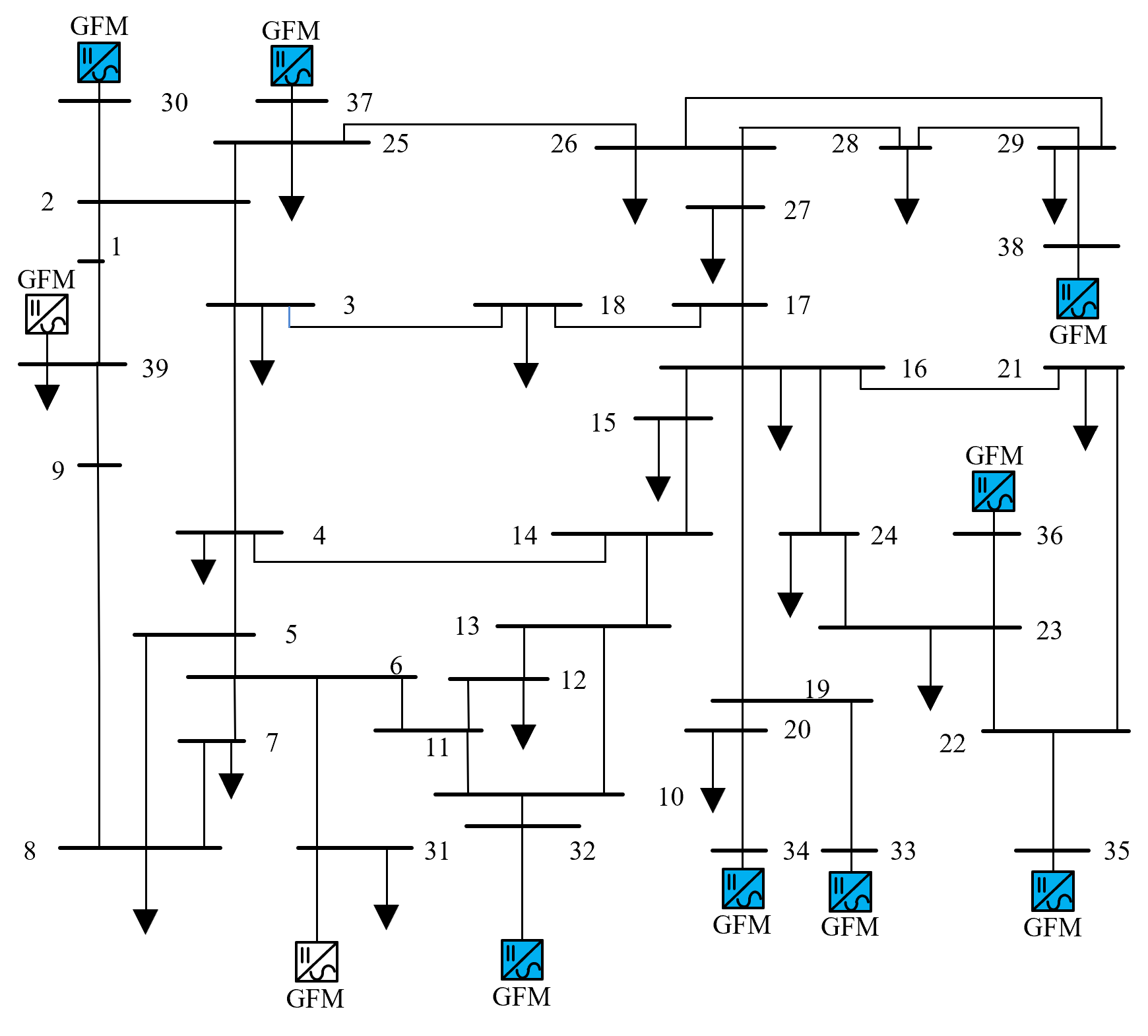}
    \caption{IEEE 39-bus test case modified for testing stability for IBR-dominated grids: every generator in the system is replaced by GFM.}
    \label{fig:IEEE39_case}
\end{figure}

This section validates the proposed method on a realistic power system model, the IEEE 39-bus test case. In its original form the system contains 10 synchronous generators; here every generator is replaced by a GFM (the GFM model is taken from \cite{pogaku2007modeling}), which emulates a possible future scenario of a 100\% renewable grid (Fig. \ref{fig:IEEE39_case}). Each GFM is rated at $200$ MVAR, and the loading of the system is scaled down accordingly. For simplicity, all transformers share the same parameters: a leakage inductance of $0.025$ pu and a resistance of $0.0025$ pu on a $200$ MVAR, $345$ kV base.

The case study uses one GFM control model, and the resulting certified droop-gain regions are specific to that model and its remaining parameters. Such a homogeneous set-up is used for simplicity in assessing the conservativeness, as the results can be shown on a 2D diagram corresponding to common voltage and frequency droop gains. The proposed methodology can nevertheless accommodate heterogeneous devices by including their distinct admittance responses as separate constraints in \eqref{eq:sdp_device} or \eqref{eq:SDP_Chebyshev_device}. Due to lack of space, simulation of multiple control architectures could not be included in this paper. However, we do simulate a heterogeneous system to some extent by varying the voltage-loop bandwidth while retaining identical droop coefficients.
 
The validation proceeds in three steps. First, we use the proposed certificate to construct a region of droop coefficients that guarantees decentralized stability. Second, we verify by direct simulation that selected droop pairs inside this region indeed yield stable systems. Third, we estimate how conservative the certified region is by searching numerically for the boundary at which stability is actually lost.
 
The direct simulations in the last two steps exploit the plug-and-play setting: we generate scenarios that differ in the placement of the $10$ GFMs across the network and assess the stability of each scenario through its eigenvalues. Because the total number of admissible placements exceeds $20$ million for this test case, an exhaustive check is out of reach; in each test we therefore sample $200$ placements at random. Such a Monte Carlo procedure can only refute plug-and-play stability, by exhibiting an unstable configuration, and can never establish it, since almost all configurations remain untested. Closing this gap is precisely the purpose of the proposed certificate.
 
\begin{figure}
    \centering
    \includegraphics[width=0.8\linewidth]{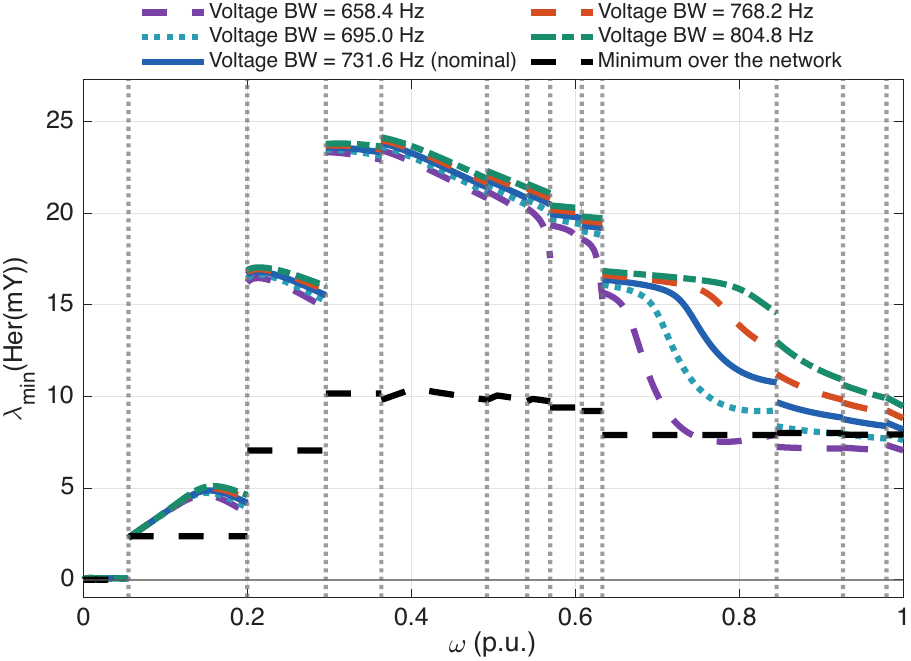}
    \caption{Minimum eigenvalues $\operatorname{Her}(mY)$ for heterogeneous GFM converters with $(m_p,n_q)=(2\%,1\%)$ and different voltage-loop bandwidths, using $3$ Hz power-filter cutoff. The black dashed curve denotes the minimum over all network lines. Positivity is retained for bandwidths of $658.4$–$804.8$ Hz.
    }
    \label{fig:eig_MY_Chebyshev}
\end{figure}

Using the full inverter model from \cite{pogaku2007modeling}, we compute a region in the $m_p$--$n_q$ plane such that, whenever the droop coefficients of every inverter lie in this region, the stability of the whole system is certified. The size of this region depends on the transformation matrices $m$ entering \eqref{eq:local_device_passivity} and \eqref{eq:local_line_passivity}, and we compare three choices: (i) two piecewise constant matrices \eqref{eq:piecewise_single}, one for each of two frequency intervals, obtained by the SDP approach; (ii) three piecewise constant matrices \eqref{eq:m_SDP_3}, one for each of three frequency intervals, obtained by the SDP approach; and (iii) a frequency-dependent matrix $m(s)$ constructed by Chebyshev approximation with Algorithm~\ref{alg:turnkey_multiplier}. In cases (i) and (ii), the frequency intervals are chosen manually, and the SDP search for the matrices is carried out at the droop pair $(m_p,n_q)=(1\%,1\%)$. In case (iii), the frequency range is \textbf{\textit{automatically}} divided into $13$ segments with provided certificate tolerance $e_H = 10^{-3}$ and maximum degree $d=10$ as inputs. All coefficients are found by a single SDP \eqref{eq:global_cheb_sdp}, so no manual splitting of the frequency range is required. The SDP search for $m(s)$ is carried out at $(m_p,n_q)=(2\%,1\%)$, a point not covered by the piecewise fixed matrices $m(s)$. The resulting minimum eigenvalues of $\operatorname{Her}(m(j\omega)Y(j\omega))$ are shown in Fig. \ref{fig:eig_MY_Chebyshev}. The methodology accommodates heterogeneous converters through their individual terminal admittances $Y_i(s)$, tested against a common multiplier $m(s)$. We illustrate this by varying the voltage-loop bandwidth while retaining identical droop coefficients in Fig. \ref{fig:eig_MY_Chebyshev}. Without multiplier reoptimization, $\lambda_{\min}(\operatorname{Her}(mY_i))$ remains positive on the evaluated frequency grid for bandwidths of $658.4$-$877.9$ Hz. This supports the applicability of the common-multiplier approach to heterogeneous converter dynamics.

\begin{figure}
    \centering
    \includegraphics[width=0.7\linewidth]{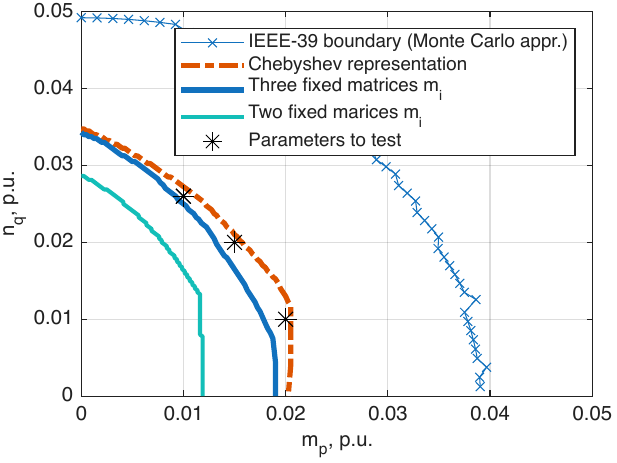}
    \caption{Plug-and-play stability regions in the $m_p$--$n_q$ plane obtained with two (solid light blue) and three (solid dark blue) piecewise constant transformation matrices $m$, and with the frequency-dependent matrix $m(s)$ obtained by Chebyshev approximation (dashed orange). Asterisks mark the droop pairs verified by direct simulation; blue crosses mark the Monte Carlo estimate of the boundary beyond which plug-and-play stability is lost. }
    \label{fig:regions_3}
\end{figure}

The certified regions for the three choices are shown in Fig. \ref{fig:regions_3}. The two SDP-optimized matrices of case (i) already yield a valid certificate (solid light blue curve), and the corresponding region is of a size that is reasonable for practical use. Splitting the frequency range into three intervals and optimizing all three matrices, case (ii), enlarges the region considerably (solid dark blue curve). Neither of these two regions, however, contains the point $(m_p,n_q)=(2\%,1\%)$. The frequency-dependent $m(s)$ of case (iii) produces the largest region of all (dashed orange curve): it covers this point and exceeds both regions obtained with piecewise constant matrices, without any manual partitioning of the frequency range. 
 
Recall the meaning of a plug-and-play stability region: the droop pair of every inverter may be chosen arbitrarily within the region, and the stability of any interconnection of any number of such inverters is then guaranteed. As a first check of the certificates in Fig. \ref{fig:regions_3}, we select three droop pairs, $(m_p,n_q)=(2\%,1\%)$, $(1.5\%,2\%)$ and $(1\%,2.6\%)$, all of which lie inside the orange region and close to its boundary (asterisks in Fig. \ref{fig:regions_3}). For each pair, we simulate $200$ random placements of the $10$ inverters, with identical droop settings on all units. In all three scenarios, every one of the $200$ scenarios is stable, consistent with the certificate.
 
As a second check, we estimate the method's conservativeness using Monte Carlo simulation. Alongside the certified regions, Fig. \ref{fig:regions_3} shows a numerically estimated outer boundary beyond which plug-and-play stability is lost (blue crosses). Each cross is obtained by fixing the voltage-to-frequency droop ratio and gradually increasing both gains until at least one of the $200$ sampled placements becomes unstable. This procedure yields an outer bound: plug-and-play stability cannot extend beyond the region enclosed by the blue crosses. The true boundary lies somewhere between the orange curve and the blue crosses, since unstable scenarios in this gap may exist that are simply not captured by the random sample. The gap between the orange curve and the blue crosses in Fig. \ref{fig:regions_3} should therefore be read as an upper bound on the conservativeness of the method; the actual gap may be smaller but cannot be determined exactly.

\section{Conclusion}
This paper has shown that decentralized stability certificates need not rely on analytical construction or heuristic guesswork. A broad class of such frequency-domain conditions shares one structure: a positive-definiteness condition that is linear in its multiplier variables. At any fixed frequency, the least conservative multiplier is therefore found by an SDP. We have demonstrated how several representative decentralized stability conditions can be written as an SDP, confirming that the approach is not tied to any single method. The optimum is guaranteed within the admitted multiplier class, and infeasibility proves that no such multiplier exists, so an unsuitable certification method is rejected automatically rather than after a fruitless manual search.

For the additive nodal-admittance formulation, we have developed a complete synthesis procedure that requires only sampled frequency responses and is therefore directly compatible with black-box EMT models and impedance scans. Component admittances and the multiplier are represented in a Chebyshev basis. The multiplier coefficients are obtained from a single joint SDP, and a certificate-driven adaptive splitting algorithm yields a piecewise-polynomial multiplier whose partition and frequency dependence within each segment are determined by the algorithm rather than by the user. The Chebyshev parameterization is not specific to this formulation: it applies to the multipliers of any certificate in the quadratically constrained class of Section~\ref{sec:other_conditions}, so that the same joint SDP and splitting procedure carry over to those methods without modification.

In the IEEE 39-bus system populated with full-order droop-controlled GFMs, SDP-optimized multipliers certify a practical region of droop gains. Direct simulation of random inverter placements confirms every tested certified setting, and a Monte Carlo estimate of the stability boundary provides an upper bound on the remaining conservatism.

Several directions remain open. 
First, some certificates may require the weight matrices to enter into nonlinear combinations, leading to non-convex optimization problems. This is the case, for instance, when the certificate is sought jointly with a coordinate transformation, so that the resulting frame is better suited to the admittances under study.
Second, since the joint SDP certifies several admittances simultaneously, it extends naturally to parametric certification. Admittances evaluated over a set of parameter values, such as operating points, loading conditions, or controller settings, can be included as additional constraints, giving, when feasible, a single multiplier valid over the entire prescribed parameter range.
% or


\appendices

\appendix \label{app:GFM_parameters}
GFM parameters for the control layout of \cite{pogaku2007modeling}:
$\omega=100\pi\,\mathrm{rad/s}$, $R=0.1\,\Omega$, $L=1.35\,\mathrm{mH}$, $\omega_c=7926\,\mathrm{rad/s}$, $K_{pc}=10.5$, $K_{ic}=16000$, $K_{pv}=0.05$, $K_{iv}=390$, $C=50\,\mathrm{\mu F}$, $f=0.75$, $R_c=0.75\,\Omega$, $L_c=36\,\mathrm{mH}$, $m_p=3.29\times10^{-4}$, $n_q=3.90\times10^{-4}$, $\omega_{\mathrm{power}}=31.41\,\mathrm{rad/s}$, $V_b=380\,\mathrm{V}$, $R_l=0.165\,\Omega$, $L_l=0.26\,\mathrm{mH}$.

% \section*{Acknowledgment}

% The Imperial College-based co-authors would like to acknowledge support by the Engineering and Physical Sciences Research Council grant number [EP/Y025946/1] and by the Leverhulme International Professorship grant reference [LIP-2020-002].
% TO DO: add ackowledgment for AEMO/CSIRO and Roy Smith for Y+E idea

\ifCLASSOPTIONcaptionsoff
  \newpage
\fi




\bibliographystyle{IEEEtran}
\bibliography{bibtex/bib/bibliography}

\end{document}